\documentclass[a4paper, 12pt]{article}
\usepackage{hyperref}
\usepackage[pdftex]{graphicx}
\usepackage{url}

\title{The micro dynamics of collective violence}

\author{Jeroen Bruggeman\thanks{Department of Sociology, University of Amsterdam, Nieuwe Achtergracht 166, 1018 WV Amsterdam, the Netherlands. Email: \texttt{j.p.bruggeman@uva.nl}. For insightful comments, thank you to Don Weenink and Frans van Winden, as well as discussants at the Violence Workshop (Amsterdam, April 2017) and the Complexity of Covert Networks workshop at the Institute of Advanced Study (Amsterdam, May 2017).}} 
\begin{document}
\maketitle

\begin{abstract}  
Collective violence in direct confrontations between two opposing groups happens in short bursts wherein small subgroups briefly attack small numbers of opponents, while the others form a non-fighting audience. The mechanism is fighters' synchronization of intentionalities during preliminary interactions, by which they feel one and overcome their fear. To explain these bursts, subgroups' small sizes and leaders' role, a social influence model and a synchronization model are compared. 
\end{abstract} 

From hunter-gatherers to modern citizens, people coalesce into groups that sometimes confront each other antagonistically. If this leads to a physical confrontation (without long range weapons), what will happen?   Randall Collins' \cite{collins08} extensive study of many photographs and surveillance video's of violent conflicts showed up the following pattern: initially, tension builds up, followed by short bursts of violence wherein small subgroups briefly attack small numbers of opponents, in particular stumbling, isolated or otherwise vulnerable individuals. Meanwhile, the others form an audience that does not commit violence \cite{weenink14}, sometimes encourages and at other times intervenes.  This overall pattern holds for both protesters and police \cite{nassauer16}, as well as groups engaging in ordinary street fights \cite{levine11}. Almost all people have to overcome a threshold to commit violence, which for most people is fear \cite{collins08}, even for police with professional training and leadership \cite{klinger04}.

People with a preference for violent action tend to select each other assortatively \cite{mcpherson01} before a fight, based on reputations and subtle signals. At a confrontation, they form small groups of typically 3 to maximally 6 individuals that include a leader. Because they are in close physical co-presence with a mutual focus of attention, their interactions strengthen their bonds \cite{sebanz06,mcneil95,collins04}. However, strong bonds are not enough to conquer fear. They have to synchronize their intentionalities \cite{tomasello07,tomasello05} through eye contact or bodily information sensed at close range. Collins \cite{collins08} emphasizes that they move rythmically and synchronize their movements, for example step back and forth together.\footnote{In lab experiments, individuals in groups moving synchronously contributed more to public goods than in other groups with asynchronous movements \cite{reddish13,fischer13}, and synchronized movements enhance feelings of strength vis-{\`a}-vis opponents \cite{fessler14}.}  Synchronization yields collective efficacy \cite{sampson97,bandura00}, called emotional energy by Collins \cite{collins08} and collective effervescence by Durkheim \cite{durkheim12}. This shared emotion makes subgroup members feel one \cite{swann12}, by which they overcome their fear. Acting violently is very exhausting, though, and lasts briefly, where after the attackers dissolve into the audience.\footnote{Highly experienced fighters do not have much fear and do not need preliminary interaction rituals to get to action. Moreover, using long range weapons or attacking unsuspecting opponents from ambush or during nightly raids is less scary than hand-to-hand combat with eye-to-eye contact, and feature different dynamics. Pre-mediated strategies, fire arms and seasoned fighters are therefore excluded from this paper.}

Current theory provides rich descriptions but leaves a number of questions unanswered. First, the temporal pattern: how does synchronization happen? Do participants synchronize in a periodic rhythm or an a-periodic pattern? Moreover, why does hand-to-hand violence occur in bursts rather than in a more gradual pattern? Second, the network: why are violent subgroups small whereas larger groups stand a better chance to gain the upper hand? Is it their small size or their high density that, through synchronization, fosters emotional energy whereas in other groups in similar circumstances, emotional energy is drained \cite{collins04}? 
Third, leadership: are leaders important for mutual alignment? Because, for example, military officers determine the choreography of marches \cite{mcneil95}, one might think they are. To answer these questions, two widely used models are applied and compared, namely social influence \cite{french56,friedkin11,friedkin16} and synchronization \cite{strogatz93,nadis03,kuramoto75,strogatz00}. Both models show how mutual alignment at the dyadic level results in synchronization at the group level, and neither of them has been applied to violence yet.

\section*{Modeling synchronization}
Following Collins, we look at situations where a small group of protagonists ($2 \leq n \leq 6$) faces one or several opponents and is surrounded by a, possibly polarized, audience. In these small groups, individuals are in close co-presence where they can clearly see, hear and sometimes touch one another. Their mutual attention is not necessarily symmetric, though. A weighted tie $a_{ij}$ (also) indicates the extent to which $i$ is receptive to $j$'s influence \cite{friedkin11}. Accordingly, the leader is the recipient of strong ties, which he reciprocates by weak ties, reflecting the asymmetry of authority versus obedience \cite{homans74}. As leaders take the initiative, their followers mimic the leader's movements with a brief delay, which informs the researcher about tie-asymmetry. 

In line with empirical findings \cite{onnela07}, the distribution of tie strengths is skewed. For practical reasons, Zipf's distribution \cite{adamic00} is used in the model, with $x$ for the rank order position of a given tie, $a_{ij} = \alpha x^{-\beta}$, and $\beta$ in the range $ 0 \leq \beta \leq  1$. Because the strongest ties are to the leader and the weakest ties from him to his followers, $\beta = 1$ indicates strong leadership, whereas $ \beta = 0$ renders all tie strengths equal and characterizes egalitarian groups. When the members focus on one another, their ties to more distant people in the larger group temporarily weaken, and are left out of the model. Now the $\alpha$ parameter can be used to express increasing tie strengths during interactions,\footnote{If not all but only some ties strengthen, their rank order changes, where after the resulting distribution is captured by the two parameters. There is a version of the synchronization model wherein tie strengths change with assortment between nodes \cite{gutierrez11}.} while egalitarian and hierarchic groups of different sizes are made comparable by making their sum totals of tie strengths identical. This is accomplished by defining normalized tie strengths as $w_{ij} = a_{ij}n/\sum_{i\neq j} a_{ij}$.


Along with ties to specific individuals, the members are committed to (identify with, or feel solidarity for) their group as a whole or its current goal, indicated by $K_i$. Because the groups' history goes beyond the scope of this paper, all members are modeled with the same level of commitment, but the model can easily incorporate distributions of commitments. As for tie strength, more strongly committed individuals are more susceptible to social influence, here of the group as a whole.

Before violence breaks out, group members differ from one another. To simplify, one key trait $\theta$ in which they differ is selected for the models, namely people's signaled intentionality, which according to Collins is rhythmic movement or speech.  Be it rhythmic or not, synchronization of signals means that their pairwise differences $(\theta_j(t) - \theta_i(t))$ stay very small over time, hence the change of signals $d\theta_i/dt$ equals zero, whereas fluctuations or large differences indicate a lack of synchronization \cite{jadba04}.\footnote{Delayed synchronization of followers in empirical data can be accommodated by the time index and incorporated in the model \cite{acebron05}.} The degree of group synchronization is indicated by an order parameter, where $r = 1$ means perfect synchronization and $r = 0$ a complete absence thereof.\footnote{To be precise, the order parameter of the synchronization model is a complex function $ r(t)e^{i \langle\theta(t)\rangle} $ \cite{strogatz00}. To enable a comparison with the influence model, an order parameter for the latter is here defined by means of normalized standard deviations, with a time index: $ r_{si}(t) = 1 - sd(\theta(t))/max(sd)$. } 

The social influence model is usually presented for discrete time steps \cite{friedkin11}, but to compare it to the synchronization model we need the continuous version \cite{abelson64}, with time indices left out for clarity, 
\begin{equation}
\frac{d\theta_i}{dt} = K \sum_{j = 1}^{n} w_{ij} (\theta_j - \theta_i), 
\label{eq:influence1}
\end{equation} 
The synchronization model \cite{kuramoto75,acebron05,rodrigues16} is:
\begin{equation}
\frac{d\theta_i}{dt} = \Omega_i + K \sum_{j = 1}^{n} w_{ij} sin(\theta_j - \theta_i), 
\label{eq:synchro1}
\end{equation} 
The latter model with the sinus term makes it possible, but does not strictly enforce, to interpret synchronization in terms of a periodic rhythm. Close to synchronization, however, the sinus drops out of the equation \cite{mcgraw07}, and therefore the salient difference with the influence model is not the rhythm but individuals' intentionalities $\Omega_i$, which in turn determine the (change of) signalled intentionalities.\footnote{Intentionalities are determined prior to the gathering by framing, prior socialization and personality, which go beyond the scope of this paper. Again for simplicity, the modeled distribution before the fight $g(\Omega)$ is symmetric around its mean. When synchronization is achieved, all $\Omega_i$ are aligned and $sd(\Omega) = 0$} The consequence of explicating them in the model is that it takes more effort in terms of social bonding, hence time, to overcome individuals' differences before synchronization is achieved.

\section*{Social dynamics}

Previous studies showed that social interactions strengthen ties \cite{sebanz06}, in particular if there is a common opponent \cite{moore79,szell10,dedreu10}. This empirical finding is represented by increasing $\alpha$ in the power law distribution, which translates into an increase of the tie strengths $w$ in the models. 

\begin{figure}
\begin{center}
\includegraphics[width=.49\textwidth]{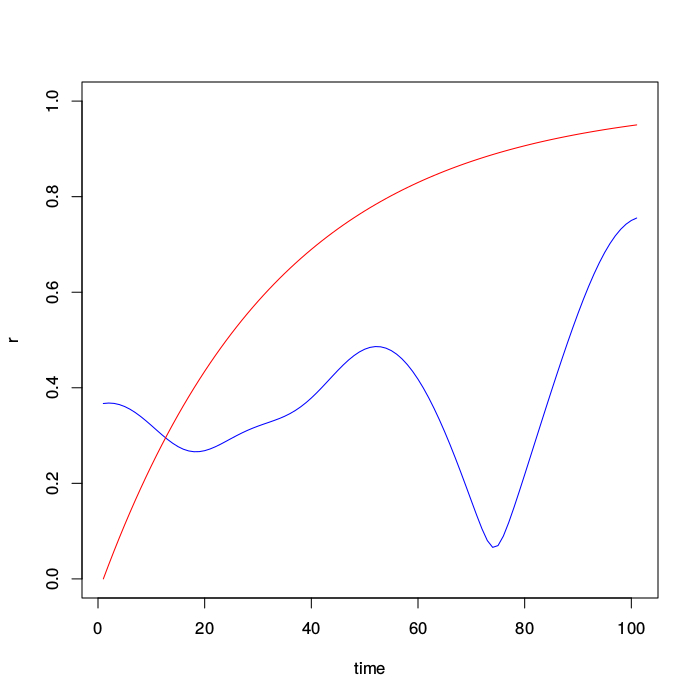} 
\includegraphics[width=.49\textwidth]{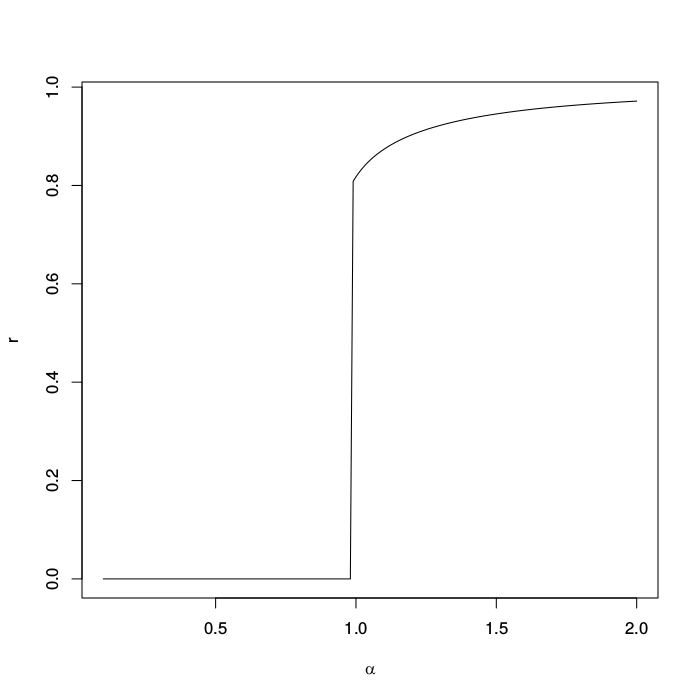} 
\caption{An egalitarian triad ($n = 3, \beta = 0$) with weak ties ($\alpha = 0.2$) at the left, and with increasing tie strengths at the right. At all $\alpha > 0$, synchronization increases smoothly over time according to the social influence model (red) but at $\alpha < \alpha_c$ it behaves chaotically according to the synchronization model (blue), which means that, strictly speaking, longitudinal synchronization is zero. The latter model is further examined at the right, now for increasing $\alpha$; at a critical threshold $\alpha_c$, stable synchronization sets in.} 
\label{increase}
\end{center}
\end{figure}

In the influence model, for every $\alpha > 0$, synchronization is realized smoothly, with the biggest increase when interactions begin. This is illustrated for a small value of $\alpha$ at the left hand side of Figure~\ref{increase}. The same figure also shows that the level of synchronization bounces around chaotically according to the synchronization model, which means that there is no synchronization at all. Only when tie strengths increase beyond a critical threshold $\alpha_c$, chaos ends and stable synchronization is achieved. In contrast to the influence model, the biggest increase of synchronization is not at the start but later on, and the tipping point is a non-intuitive result of this model \cite{arenas08}.  Synchronization increases collective effervescence that can be used for an attack. The sociological application of the models has an additional feature: synchronization also increases solidarity ($K$) \cite{collins04} that, through a feedback loop, enhances collective effervescence. 

When the focal group is synchronized it becomes \emph{critical} \cite{daniels17}. This means that the group becomes exceptionally sensitive to minor provocations or ephemeral vulnerabilities of the opponent(s), which then trigger the group into violent action. When the group is still asynchronous, and at a distance from criticality, such small events may heat up emotions but do not trigger a collective attack. Criticality explains why violence occurs in bursts. When in the course of action the fighters get exhausted, their attack unravels. 

Online videos of street violence are truncated and miss the beginning. Therefore it is not possible at this point to favor one model over the other. In these videos, synchronized individuals show up 
a---non rhythmic---alternation of movements such as moving forward and backward, possibly because a lack of periodicity makes them less predictable for opponents. These alternations can be represented by the synchronization model as different phases of movement, but because there is no periodic rhythm, this argument is not decisive.  
 
\begin{figure}
\begin{center}
\includegraphics[width=\textwidth]{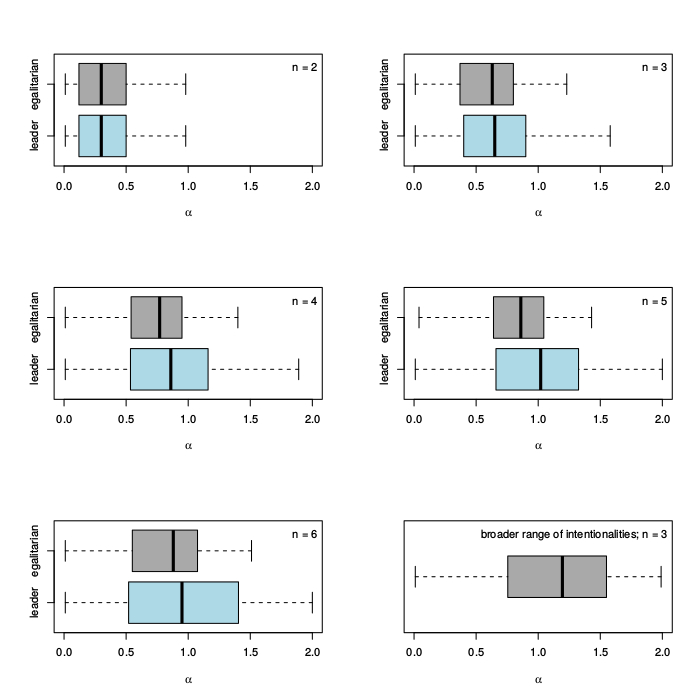} 
\end{center}
\caption{Distributions  of  tipping  points.} 
\label{boxplot}
\end{figure}

Let us now examine small and fully connected networks ($2 \leq n \leq 6$). For each of these networks, the synchronization model is solved numerically, by randomly drawing sets of intentionalities and initial values of signals 500 times, and implementing them in both the leadership ($\beta = 1$) and egalitarian versions ($\beta = 0$). The distributions of the tipping points are shown in Figure~\ref{boxplot}.  As the widths of the boxplots show, the tipping points are sensitive to the variability of intentionalities and initial values; if by chance, or assortative selection by the fighters, these values are more similar, stable synchronization is achieved at lower tie strengths. To emphasize this fact, the egalitarian triad is depicted once more (at the bottom) with a range of intentionalities twice as broad as the other boxplots. 

Figure~\ref{boxplot} also shows that leaders do not facilitate synchronization. This counter-intuitive result can not be obtained empirically, because one will confound leadership and tie strengths. In the model, in contrast, overall tie strengths of different groups under comparison can be kept exactly the same. For a leader to have his group synchronize faster, there should thus be at least two factors at play: a skewed tie strength distribution and stronger ties to him (or more similar individuals to begin with). 

Furthermore, because people have cognitive limitations, they can not pay attention to everybody else in large groups. Large networks are therefore sparser than small cliques, typically have longer distances, and possibly topological bottlenecks where different clusters are connected by a single tie or node. Different network topologies and tie strength distributions can be compared by their {\em algebraic connectivity} \cite{jadba04},\footnote{Algebraic connectivity is the second smallest eigenvalue of the network Laplacian \cite{fiedler73,abreu07}. The Laplacian $\mathcal{L}$ of an asymmetric matrix $\bf W$ with weighted ties, and ${\bf P}$ the row normalized matrix thereof, is defined as 
${\bf \mathcal{L}} = {\bf I} - \frac{ \Phi^{1/2} {\bf P} \Phi^{-1/2} + \Phi^{-1/2} {\bf P}^T \Phi^{1/2} }{2}$,
where $\bf{I}$ is the identity matrix, and $\Phi$ the Perron vector of ${\bf P}$ written as a diagonal matrix \cite{chung05}. For cliques, each diagonal cell in $\Phi$ equals $1/n$; in general these cells can be assessed by weighted PageRank \cite{prystowsky05}. } denoted $\lambda_2$.  If $\lambda_2 = 0$, the network is disconnected and won't synchronize, whereas it is maximal in maximally connected networks, which synchronize most easily (at small $\alpha$). In egalitarian networks, the maxima are $\lambda_2 = 2$ in a dyad, $\lambda_2 = 1.5$ in a triad, and they decrease asymptotically to 1 with increasing network size (and keeping density maximal), which corresponds to the increase of (mean) tipping points in Figure~\ref{boxplot}. Skewing the tie strength distribution lowers algebraic connectivity (if $n > 2$) and pushes the mean tipping points in Figure~\ref{boxplot} slightly to the right, in line with the numerical results above. For networks with unfavorable topologies to synchronize, e.g.~when some people can not see each other, leaving structural holes in the network, very strong ties (or commitments) are required to compensate \cite{jadba04,dorfler14}, unless intentionalities are very similar at the beginning. This explains why bursts of violence are committed by fully connected {\em small} groups, even though larger groups would have a better chance to beat their opponents.\footnote{If, alternatively, the effect of increasing $K$ is modeled in large networks \cite{arenas06}, it turns out that synchronization starts in relatively well-connected clusters that are much smaller than the overall network, consistent with the argument made here.}   This is a third non-obvious result, and the second one that can not be obtained by empirical research. Because in the social influence model, the speed to synchronization also depends on the algebraic connectivity of the pertaining networks \cite{olfati07}, all results discussed here hold true for it, except the tipping points.

\section*{Discussion}  

The modeling of pre-fight interactions yielded a threefold contribution to the discursive theory of violence and thereby increases its explanatory power. First, increasing tie strengths during interactions in close co-presence result in synchronization, which yields collective efficacy and renders the group critical. Then, a minor insult or a tripping opponent can easily trigger an attack. According to Collins \cite{collins04}, the movements during preliminary interactions are rhythmic, and rhythm is certainly important in ``dance and drill" \cite{mcneil95} and in many other interaction rituals \cite{collins04}. Video's of street violence, however, show that there is no rhythmic entrainment during preludes to street fights, but a-periodic alternations of movements. These alternations suggests that the synchronization model (with subsequent phases) is better suited than the influence model, but the lack of periodicity leaves some doubt, which remains unresolved because the online videos are left-truncated. The next version of this paper will be co-authored by Don Weenink and incorporate an analysis of 50 video's coded by him, in contrast to the cursory look that I had at some of them. This material will help to get a better understanding of the temporal pattern of street violence.



Second, tie strengths together with network topology---summarized by algebraic connectivity---have to compensate for individuals' heterogeneity \cite{dorfler14}. When algebraic connectivity is low, collective effervescence will stay low no matter how hard individuals try their best. This explains why violent groups are typically small, and helps to explain Collins' \cite{collins04} observations that sometimes emotional energy is drained. Now one might think that at large open spaces, for example Tahrir square in Cairo, visual cues make it possible for large numbers of protesters to align signals, like holding up shoes. Although this is true, long range visual ties across big spaces are weak, and long range synchronization inaccurate. In physical confrontations with the police, only small groups engaged, in line with the models' prediction. 

Third, the models shows that without sheer luck, leaders can not speed up synchronization faster than  non-hierarchic groups. The videos show that their main role is to take the \textit{initiative} in motility and attack, and possibly in organization before the camera is on \cite{glowacki15a}. In other situations, leaders may use choreography to synchronize their groups, seen in military marches throughout the world. Without eye contact, however, the effect of rhythmic movement is not strong enough \cite{mcneil95}, and soldiers' preparations for violence require more than marches. 

The three main findings are counter-intuitive. In the age of big data it is important to stress that without a good theoretical---versus statistical---model, many insights can not be obtained, no matter how big one's data is. Here, the second and third result could only be had by means of a model that disentangles effects (tie strength and leadership) that can not be kept apart in field studies. The three findings seem to be robust because they follow from both influence and synchronization models. 

Currently it is not completely understood why small numbers of people are willing to take disproportional high risks for their larger group's victory. Combatants are aware that they can improve their reputations in terms of attributed status or prestige \cite{henrich01,homans74}, corroborated by more severe violence under audience presence, which partly explains their motivation. They are also motivated by an enhanced sense of belonging \cite{whitehouse14b,baumeister95}. It is not clear, however, if the benefits exceed the costs. Moreover, people may satisfy their need to belong in different ways, for example by dancing synchronously and thereby precluding injury. Future studies could investigate to what extent short term motives align with long term cost-to-benefit ratios \cite{glowacki15b}, for example by expanding the model with a group's broader network embedding and history \cite{papachristos09,glowacki16}, which seem also important to predict the severity of violence. 





\small


\end{document}